\documentclass[a4paper]{PoS}

\usepackage{IEEEtrantools}

\title{The lateral distribution function of cosmic-ray induced air showers studied with the HAWC observatory}

\ShortTitle{LDF of EAS measured at HAWC}

\author{\speaker{J. A. Morales-Soto}, J. C. Arteaga-Vel\'{a}zquez and J. D. \'{A}lvarez for the HAWC collaboration\thanks{For the collaboration list see PoS(ICRC2019)1177. Complete list of authors at https://www.hawc-observatory.org/collaboration/icrc2019.php }\\
        Instituto de F\'{i}sica y Matem\'{a}ticas, Universidad Michoacana de San Nicol\'{a}s de Hidalgo, Mexico\\
        E-mail: \email{jmoralessg@ifm.umich.mx}}

\abstract{The particle lateral distribution function (LDF) of air showers at a given altitude is sensitive to the mass composition and primary energy of cosmic rays. Studies of the LDF are difficult to perform due to experimental effects such as sampling bias, as well as shower-to-shower fluctuations in particle density. The High Altitude Water Cherenkov (HAWC) observatory, a dense air shower array located in central Mexico at 4100 m a.s.l., is well-suited to perform detailed event-by-event studies of the LDF of multi-TeV cosmic-ray showers. The detector is instrumented with 1,200 photomultipliers (PMTs) in close-packed water Cherenkov tanks containing a total of 60 ML of water. We present a study of the LDF of cosmic-ray air showers recorded by HAWC in 2016 with energies between 3 TeV and 300 TeV and zenith angle < 16.7$^{\circ}$. The data are used to determine the optimal parameterization of the LDF at the HAWC site. From here the lateral shower age is obtained and its sensitivity to the cosmic ray mass composition is analyzed.}

\FullConference{36th International Cosmic Ray Conference -ICRC2019-\\
		July 24th - August 1st, 2019\\
		Madison, WI, U.S.A.}

\begin{document}

\section{Introduction}

At high energies (E > 100 TeV) the statistics of cosmic-ray measurements performed by instruments placed at the top or outside our atmosphere (such as air balloons or satellites) decreases due to the fact that the flux of this radiation decays following a power law. Due to this fact it is necessary to perform indirect measurements through extensive air shower (EAS) detectors installed at ground level, which can compensate the small flux of cosmic rays at energies > 100 TeV thanks to their big size.

Over the years, a variety of instruments have been developed with the purpose of studying the EAS on Earth's surface. The detailed study of the characteristics of the air showers (such as lateral structure, size, age, particle content, fluorescence emission, etc.) provides information about the properties of the primary cosmic and gamma rays of high energies \cite{Grieder}.

For example, the age parameter, \emph{s}, which gives a description of the development stage of the air shower in the atmosphere, has a dependence to the mass composition of the cosmic rays. The concept of age parameter was born from the study of longitudinal and lateral development of electromagnetic showers  generated by photons and electrons \cite{NKG_ref}. Even though this concept was developed for electromagnetic EAS, it can also be extended to air showers of hadronic nature. 

The longitudinal age parameter depends on the production, energy spectrum  and decay of secondary particles in the shower. In the case of an electromagnetic shower, the longitudinal age parameter is related with the atmospheric depth, $X$, by \cite{age_longitudinal}:

\begin{equation}
    s = \frac{3t}{t + 2\beta},
\end{equation}

\noindent with $\beta$ = ln(E/E$_{c}$) and $t=X/X_{0}$, where $E_c$ is the critical energy for the production of secondary particles, $X_{0}$ = 1030 g/cm$^{2}$ is the atmospheric depth at sea level.

On the other hand, the lateral age parameter is related to the lateral distribution of particles around the shower axis. In particular, the value of the lateral age indicates the slope of the lateral distribution \cite{NKG_ref}. Showers that were generated high in the atmosphere are called \emph{old}, and are characterized  by having large $s$ values. In general, they are produced by low energy events and heavy primaries. \emph{Young} air showers are generated at lower altitudes in the atmosphere, posses smaller values of lateral age and are related to high energy events and light primaries. For a pure electromagnetic cascade, the Lateral Distribution Function (LDF) and the lateral age parameter, $s$, are related through a Nishimura-Kamata-Griesen (NKG) function \cite{NKG_ref} 

\begin{equation}
    f(r) = N_e C \left( \frac{r}{r_M} \right)^{s-\alpha} \left( 1 + \frac{r}{r_M} \right)^{s-\beta},
    \label{Eq:lateral age}
\end{equation}

\noindent where $N_e$ is the number of electrons at the observation level, $C$ is a normalization factor, $r_M$ is the Moliere radius and $\alpha$ = 2 and $\beta$ = 4.5 are fixed parameters.

Frequently, the NKG function, or a modified version of it, is used for the description of the lateral density distribution of electrons and other secondary particles, such as muons \cite{KASCADE_LDF1} in EAS generated by cosmic rays. 

The interest in the study of the lateral distribution lies in the fact that it contains information about the energy and the mass composition of the primary cosmic rays.

In this work, the lateral distributions of cosmic-ray induced showers recorded by HAWC in 2016 in the energy interval E = 10$^{3.5}$ GeV - 10$^{5.5}$ GeV and zenith angles $\theta$ < 16$^{\circ}$ are studied with the aim to find their optimal parametrization. From here the lateral shower age is obtained and its sensitivity to the cosmic ray mass composition is analyzed.

HAWC is an air shower detector designed to study the sky in gamma rays (E = 100 GeV - 100 TeV), and cosmic rays (up to 1 PeV). The experiment is located at 4100 m a.s.l. at the Pico de Orizaba Volcano in Puebla, Mexico and consists on an array of 300 water Cherenkov detectors that covers 62$\%$ of a flat surface of 22,000 m$^2$. HAWC is instrumented with 1,200 photomultipliers (PMT) and 60 ML of water. The PMTs record the arrival time of the secondary particles as well as the lateral distribution of the deposited charge of the EAS. The HAWC observatory is well-suited to perform detailed event-by-event studies of the lateral distribution of EAS at TeV energies given its instrumented area, large volume and physical coverage, closed-packed detector design, high altitude and large number of PMTs. 

\section{HAWC data and simulation}

The experimental data set used for this work covers the period from 2/06/2016 to 12/06/2016 and has a total duration of 1.3 days. To diminish the systematic errors in the analysis, the events must fulfill some quality criteria: they must have successfully passed the event reconstruction procedure described in \cite{CRAB_ref}, their arrival direction must have $\theta$ <16.71$^{\circ}$, they should have at least 60 activated PMTs in a radius of 40 m from the shower core, their core must be inside HAWC's area, there must be at least 75 hit PMT's, they must have at least 30$\%$ of the available channels with signals. After applying the selection criteria, the experimental data set has a total of 21,922,210 events. The cuts used for the experimental data set were studied with Monte Carlo (MC) simulations. To simulate the creation and propagation of air showers in the atmosphere CORSIKA (v740) \cite{CORSIKAref} was employed using the hadronic interaction models \emph{FLUKA} \cite{Ferrari2011}, and \emph{QGSJet-II-03} \cite{Ostapchenko2005}, for low and high energies respectively, while the interaction of the EAS  with HAWC's detectors were simulated with a code based on \emph{GEANT4} \cite{Agostinelli2003}. Protons and iron nuclei were simulated using the spectrum described in \cite{Light_ref} MC simulations were reconstructed performed using the same reconstruction algorithm applied to the experimental data. The primary energy reconstruction method is described at \cite{Light_ref}. Also, the experimental data set is in the energy interval between E = 10$^{3.5}$ GeV - 10$^{5.5}$ GeV.

\section{Study of different Lateral Distribution Functions}

At HAWC, the lateral distribution of an event, $Q_{eff}(r)$, is reconstructed by plotting the effective charge distribution deposited at the PMTs of HAWC against the lateral distance to the shower core, using the shower front coordinate system. The errors of the effective charge are assigned by using the following empirical function:

\begin{equation}
\log_{10}(Q_{error}) = \left\{ \,
\begin{IEEEeqnarraybox}[][c]{l?s}
\IEEEstrut
0.3-0.06667\cdot \log_{10}(\mathnormal{Q}_{eff}) \,\,\,\,\,\,\,\,\, & if \, $\log_{10}(Q_{eff})\leq$3, \\
0.1 & if \, $\log_{10}(Q_{eff})>$3. 
\IEEEstrut
\end{IEEEeqnarraybox}
\right.
\label{eq:example_left_right1}
\end{equation}

An example of a lateral distribution of an event measured at HAWC is shown in Fig. \ref{Fig:sample} (left).

During the analysis, the lateral distribution of an EAS event is fitted using a modified NKG function \cite{CRAB_ref}:

\begin{equation}
        f(r) = A \left( \frac{r}{r_M} \right)^{s-3} \left(1 +  \frac{r}{r_M} \right)^{s-4.5},
    \label{Eq:LDFHAWC}
\end{equation}

\noindent where $r_M$ = 124.21 m is the Moliere radius at HAWC altitude \cite{CRAB_ref}, A is a normalizing factor, and $s$ is the lateral age parameter. Here $A$ and $s$ are free parameters. It has been proven that this NKG-like LDF gives a good description of gamma-ray induced air showers detected with HAWC \cite{HAWCLDF_gamma}, but this hasn't been proven yet for cosmic ray induced air showers. For this reason, in this work, this and other LDFs were chosen from the literature and were fitted to the data in order to find a LDF that best describes the lateral distribution of EAS created by cosmic rays and detected with HAWC.

Besides the function (\ref{Eq:LDFHAWC}), the second function from the list of selected LDFs is modified NKG-function used by the KASCADE collaboration to fit their shower data \cite{KASCADE_LDF1,KASCADE_LDF2}:

\begin{equation}
    f(r)=N \cdot \bar{c}(s) \cdot \left( \frac{r}{r_0} \right)^{s-\alpha} \left(1 + \frac{r}{r_0} \right)^{s-\beta},
    \label{Eq:NKG_KASCADE}
\end{equation}

\noindent where $N$ a normalization factor, $r_0$ is a scale parameter, and 

\begin{equation}
    \bar{c}(s) = \frac{\Gamma(\beta - s)}{2 \pi r_{0}^{2} \Gamma(s -\alpha  +2) \Gamma(\alpha + \beta -2s -2)}. 
\end{equation}

This function was applied individually to the electron, muon, and hadron lateral distributions in the energy range E = 100 TeV - 1 EeV.

On a similar study of the lateral distribution of EAS, the ARGO-YBJ collaboration found that the LDF that best described it's simulated and experimental data was the following NKG-like function \cite{ARGOLDF} :

\begin{equation}
    f(r)=A\left( \frac{r}{r_0} \right)^{s-2} \left(1 + \frac{r}{r_0} \right)^{s-4.5},
    \label{Eq:Argo_LDF}
\end{equation}

\noindent where $A$ is a normalization factor, $s$ is the lateral age parameter, and $r_0$ = 30 m is a constant scale radius. This function was applied individually to simulations and measured data in the energy range E = 50 TeV - 10 PeV. 

Finally, the AGASA and ARGO collaborations also employed the following LDF called the \emph{scaling formalism} to fit their data on EAS \cite{AGASA,ARGOLDF}

\begin{equation}
        f(r) = C \left( \frac{r}{r_0} \right)^{-\alpha} \left(1 +  \frac{r}{r_0} \right)^{-(\beta-\alpha)} \left[1+ \frac{r}{10r_0} \right]^{\delta},
    \label{Eq:LDFScaling}
\end{equation}

\noindent where C is a constant, $\alpha$, $\beta$ and $\delta$ are fixed parameters, while $r_0$ is a scale radius left as a free parameter that is shown to be correlated with the shower age. A modified version of Eq. (\ref{Eq:LDFScaling}) was selected for this work

\begin{equation}
        f(r) = C \left( \frac{r}{r_M} \right)^{-s} \left(1 +  \frac{r}{r_M} \right)^{(s-\beta)} \left[1+ \left( \frac{r}{r_M}\right)^{\phi} \right]^{\delta},
    \label{Eq:LDFModified}
\end{equation}

\noindent where $\beta$ = 3.59, $\phi$ = 0.19 and $\delta$ = 3.61 were derived from MC simulations and $r_M$ = 124.21 m is the Moliere radius. In the following,  Eq. (\ref{Eq:LDFModified}) will be given the name of \emph{modified scaling formalism}.

\section{Analysis and results}

\textbf{Fits of the LDFs}

 To fit the experimental data the previous LDFs were employed using the $\chi^{2}$ method to verify the quality of the fits. 

\begin{figure}
	\begin{center}
		
		\hspace*{-0.3cm}\begin{tabular}{ c c }
		
		\includegraphics[width=0.5\linewidth,height=0.45 \linewidth]{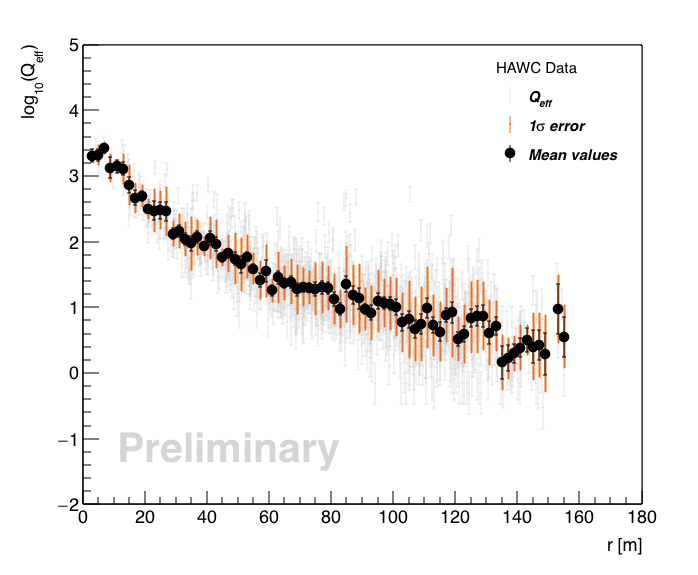}  &
		\includegraphics[width=0.4\linewidth,height=0.43 \linewidth]{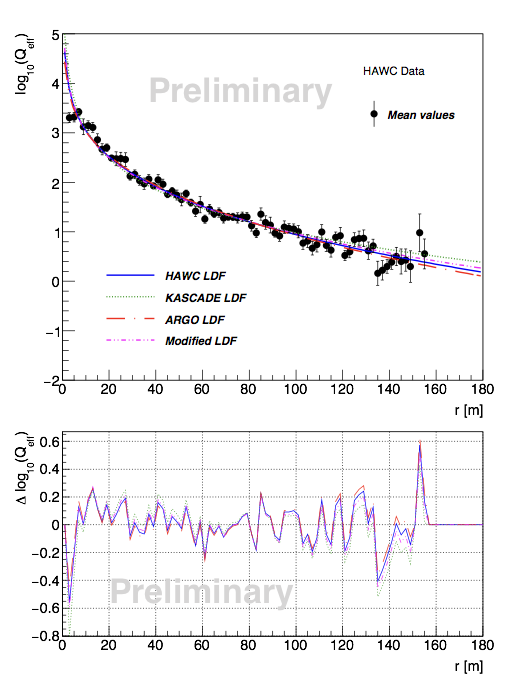}
		
	\end{tabular}
	\end{center}
	\caption{Left panel: lateral distribution of an air shower from the experimental data set of an energy of $E=10^5$ GeV and a zenith angle of $1.3^{\circ}$. The gray markers correspond to the effective charge points per PMT and their corresponding errors were calculated using Eq (\ref{eq:example_left_right1}), the solid black circles correspond to the mean $Q_{eff}$ per radial bin and there the error bars represent the error on the mean and the sigma error. Right panel: result of the fits to the mean $Q_{eff}$ (top). The difference between the observed values and the values from the fit (bottom). The $\chi^{2}/NDOF$ values of the results from the fits are: HAWC LDF = 1.69, KASCADE = 2.54, ARGO LDF = 1.52, Modified LDF = 1.85.} 
	\label{Fig:sample}
\end{figure}

First the mean of the deposited effective charged deposited, $\bar{Q}_{eff}$, is obtained for several radial bins, $r$ of size 2 m, and then different fits with the parameterizations described above are applied over the $\bar{Q}_{eff}$ distributions using the $\chi^{2}$ method (see Fig. \ref{Fig:sample} right).

To obtain the LDF that best fits the data,  a systematic study was carried out through a comparison of the $\chi^{2}$ per number of degrees of freedom (NDOF) of the results of the fits. 

\begin{figure}
	\begin{center}
		
		\hspace*{-0.3cm}\begin{tabular}{ c c }
		
		\includegraphics[width=0.5\linewidth,height=0.45 \linewidth]{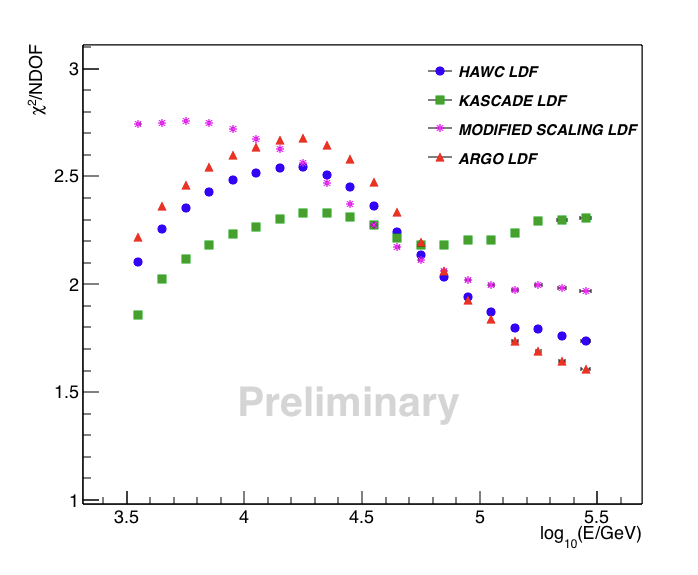}  &
		\includegraphics[width=0.5\linewidth,height=0.45 \linewidth]{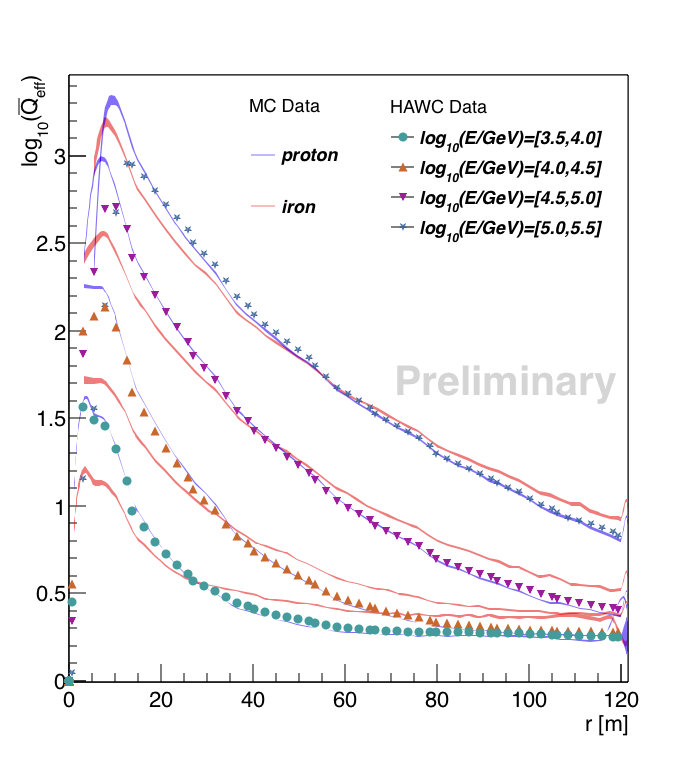}
		
	\end{tabular}
	\end{center}
	\caption{Left panel: comparison $\chi^{2}/NDOF$ of the resulting fits of each of the selected LDFs for the experimental data. Right panel: Average lateral distribution of an EAS for different energy intervals corresponding to the MC (blue bands: proton; red bands: iron) and experimental (solid markers) data. In both plots, the errors on the mean are smaller than the marker size.} 
	\label{Fig:chi_and_age}
\end{figure}

The mean $\chi^{2}/NDOF$ of the fits to the data as a function of the energy for the interval from 10$^{3.5}$ GeV to 10$^{5.5}$ GeV are shown in Fig. \ref{Fig:chi_and_age}.

The average $Q_{eff}(r)$ of an EAS for different energy intervals of the MC and experimental data were estimated to test the predictions of the high energy hadronic interaction model \emph{QGSJet-II-03} (see Fig. \ref{Fig:chi_and_age}).

\textbf{Lateral age parameter sensitivity to the mass composition}

\begin{figure}
	\begin{center}
		
		\hspace*{-0.4cm}\begin{tabular}{ c c }
		
		\includegraphics[width=0.5\linewidth,height=0.45 \linewidth]{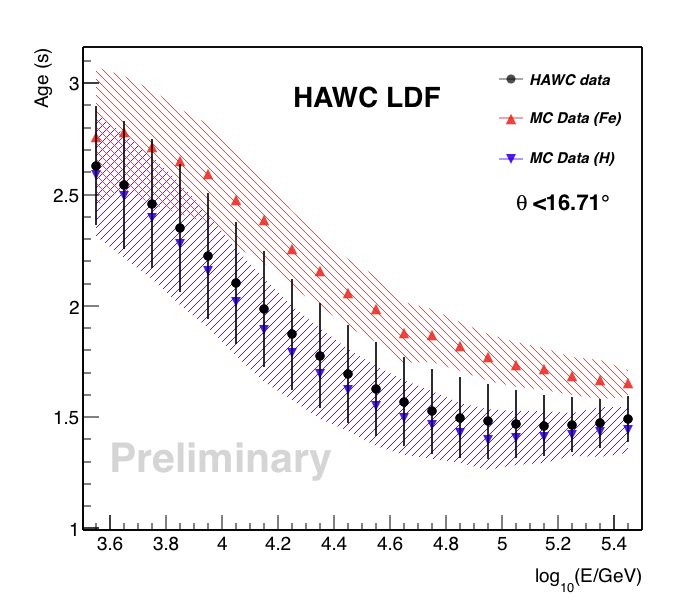}  &
		\includegraphics[width=0.5\linewidth,height=0.45 \linewidth]{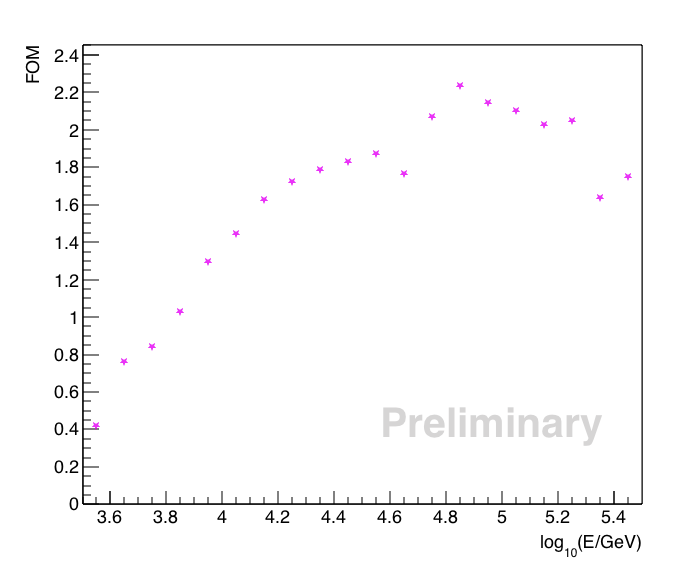}
		
	\end{tabular}
	\end{center}
	\caption{Left panel: lateral age parameter estimated by the fit performed from the HAWC LDF (\ref{Eq:LDFHAWC}) as a function of the energy. The error bands of each component correspond to a 1$\sigma$ containment error. Right panel: FOM distribution for the proton and iron calculated with MC data.} 
	\label{Fig:mean_curves}
\end{figure}

To study the sensitivity of the $s$ parameter to the cosmic ray composition, we have analyzed the average $s$ obtained with the fit using the HAWC LDF with the MC and experimental data, which is shown in Fig. \ref{Fig:mean_curves} (left). In the plot, the error bars (bands) represent the one sigma error. It can be seen that the age parameter has a dependence with the mas composition and the energy. For a quantitative description of the sensitivity of the age parameter the figure of merit (FOM) was evaluated. The FOM quantifies the separation between the mean values classes to analyze \cite{FOM_ref}, in this case, the separation of proton and iron showers by the age value. The FOM is defined as: 
\vspace{-0.3cm}
\begin{equation}
    FOM = \frac{|s_{Fe}-s_{p}|}{\sqrt{\sigma_{p}^{2}+\sigma_{Fe}^{2}}},
\end{equation}
\vspace{-0.3cm}

\noindent where $s_p$ and $s_{Fe}$ are the mean values of the proton and iron populations, respectively, and $\sigma_{p}$ and $\sigma_{Fe}$ are the one standard deviation of the proton and iron classes. The FOM defines a relation between the mean values and the standard deviations of the observable for the two populations thus, for example, if the FOM = 1 it indicates that the means of the two populations are separated by one standard deviation. The FOM for the age MC distributions shown in Fig. \ref{Fig:mean_curves} (left) are shown on the plot of Fig. \ref{Fig:mean_curves} (right).

\section{Comparison of LDF data with model predictions}

To test the interaction models we have compared the MC predictions with the measured data for the mean lateral distribution at EAS at four different energy intervals from E = 10$^{3.5}$ GeV to 10$^{5.5}$ GeV. The results are presented on Fig. \ref{Fig:chi_and_age} (right) and discussed in the following section.

\section{Discussion}
From Fig. \ref{Fig:chi_and_age} (left), for energies E < 10$^{4.5}$ GeV the Eqs. (\ref{Eq:NKG_KASCADE}) and (\ref{Eq:LDFHAWC}) give a better description of the data, but in the high energy regime, E > 10$^{4.5}$ GeV, Eqs. (\ref{Eq:Argo_LDF}) and (\ref{Eq:LDFHAWC}) fit better to the data. That means that the HAWC LDF gives also a good description of cosmic-ray induced EAS.

By studying Fig. \ref{Fig:mean_curves} (right), it can be seen that the value of the FOM for the proton and iron populations is greater than 1 for energies E > 10$^{3.8}$ GeV and reaches its maximum at E = 10$^{4.8}$ GeV, and then it starts to decrease. In particular, at E = 10$^{5.5}$ GeV the FOM is equal to 1.75.

\section{Conclusions}

As shown in the left panel of Fig. \ref{Fig:chi_and_age}, it has been found that none of the selected LDFs describes satisfactorily the measured data for all radial ranges and energies. However, it was found that HAWC's LDF gives a good description of the data in the energy range between E = 10$^{3.5}$ - 10$^{5.5}$ GeV in comparison with the other LDFs. On the other hand, from the results of the FOM it can be ensured that it is possible to perform mass composition studies with HAWC's data in an energy interval between E = 10$^{3.8}$ - 10$^{5.5}$ GeV using the lateral age parameter. Finally, from Fig. \ref{Fig:chi_and_age} (right) it can be seen that the average $Q_{eff}(r)$ of an EAS follows within the predictions of the hadronic interaction model QGSJET-II-03.

\acknowledgments

The complete information about the acknowledgments can be found at https://www.hawc-observatory.org/collaboration/icrc2019.php

Also, J. A. Morales-Soto and J. C. Arteaga-Vel\'{a}zquez would like to thank CONACYT grant A1-S-46288.

\end{document}